\documentclass[prb]{revtex4}
\usepackage{graphicx}
\usepackage{dcolumn}
\usepackage{bm}

\begin{document}

\title{\bf Weak antilocalization in a strained InGaAs/InP quantum well structure}
\author{S.A. Studenikin\footnote{sergei.studenikin@nrc.ca}, P. T. Coleridge, and P. J. Poole}
\affiliation{Institute for Microstructural Sciences, National
Research Council, Ottawa, Ontario, K1A OR6, Canada}

\begin{abstract}
Weak antilocalization (WAL) effect due to the interference corrections to the conductivity  has been studied experimentally in a strained InGaAs/InP quantum well structure. 
From measurements in tilted magnetic filed, it was shown that both weak localization and WAL features depend only on the normal component of the magnetic field for tilt angles less than 84 degrees.   
Weak antilocalization effect showed non-monotonous dependence on the gate voltage which could not be explained by either Rashba or Dresselhouse mechanisms of the spin-orbit coupling. 
To describe magnetic field dependence of the conductivity, it was necessary to assume that spin-orbit scattering time depends on the external magnetic field which quenches the spin precession around effective, spin-orbit related, magnetic fields.
\end{abstract}


\keywords{weak localization, antilocalization, phase-breaking time, spin-orbit coupling, magnetoresistance, magnetotransport}


\maketitle

\section{Introduction}
When spin-orbit scattering is strong the weak localization
feature in a 2-dimensional system develops an antilocalization structure \cite{Altshuler85,Hikami80}. 
This appears as an additional negative magneto-resistance at very low fields and
provides a convenient means of monitoring the spin-orbit
interaction \cite{Koga02,Miller03,StudPRB}  an understanding of which is
needed if spins are to be manipulated for spintronic and
quantum computing applications in semiconductors \cite{Wolf01}.
Results are presented here for a strained InGaAs/InP quantum
well, where the antilocalization feature is prominent, but which
cannot be explained using currently available theories.

\section{Experimental}
The QW structure, grown on a (100) InP semi-insulated
substrate, consisted of: 450 nm of undoped InP buffer layer, 10
nm In$_{x}$Ga$_{1-x}$As (with {\it x}=0.76), a 13 nm undoped InP
spacer layer,  13 nm of InP doped with Si at $4\times10^{17}$cm$^{-3}$) 
and a 13 nm undoped cap layer. Because the lattice
constant of InP is 1.53\% less than 76\% InGaAs the quantum well
is compressively strained in the plane.  A gated Hall bar was
prepared using standard optical lithography and wet etching
techniques.  At zero gate voltage the concentration was
4.9$\times$10$^{15}$m$^{-2}$ and the electron mobility 7.8
m$^2$/Vs.  Measurements were made in a He3 system with a split
tranverse axis superconducting coil.

Figure 1 shows typical magnetoresistance traces at T=0.35 K for
various magnetic fields tilted by the angles shown away from the normal to the
surface. When plotted against R$_{xy}$ (which depends only on
the normal component of the magnetic field ) it can be seen
[Fig.1(b)] that the data collapses onto a single curve. That is (at
least up to tilt angles of 85$^0$) both the weak localization
and weak antilocalization (WAL) features depend only on the
perpendicular component of the magnetic field and therefore result from  
orbital motion. This is in contrast to the intuitive
expectation that the antilocalization depends on spin and that
tilting the magnetic field should decouple the spins from the
orbital motion.  Similar results have also been observed in an
isomorphous (unstrained) sample \cite{StudPRB}.

Figure 2 shows magnetoconductivity traces for various gate
voltages (V$_g$) between +0.1 V (corresponding to a density of
5.4$\times$10$^{15}$m$^{-2}$) and -0.6V (1.4$\times$10$^{15}$m$^{-2}$) 
A feature of these results is that the WAL peak 
depends non-monotonically on gate voltage: it
is absent at positive gate voltages,  increases with decreasing
V$_g$ to reach a maximum at -0.3V and then decrease again with
further decreases in gate voltage.  It is qualitatively evident
from Fig. 2 that the spin-orbit scattering parameter 
$\beta_{so}$ (defined as $ \tau/\tau_{so}$ where $\tau$ and $\tau_{so}$ are 
respectively the elastic and spin-orbit scattering times) depends non-monotonically 
on the gate voltage. The maximum at -0.3V corresponds to a density of
3.4$\times$10$^{15}$m$^{-2}$. This behaviour cannot 
be explained by either the Rasba or the Dresselhaus mechanisms \cite{Miller03,Dresselhaus92,StudJETP},
both of which predict a monotonic dependence on 
density.

\section{Discussion}
The solid lines in Fig. 2 are attempts to fit the data using a recent
theory for the weak localization effect valid in arbitrary magnetic
fields \cite{Gasparyan85,Zduniak97}. The data in Fig. 2 are plotted
 vs normalized magnetic field $B/B_{tr}$,where 
$B_{tr} = \hbar/4eD\tau$  is  the characteristic transport field. The theory
assumes a single spin-orbit scattering mechanism dominates. For InGaAs structures 
this is expected to be the Rasba term which should be significantly larger than 
the bulk (Dresselhaus) term. It should be noted that  at low magnetic fields 
this theory coincides with the exact analytical solution
due to Hikami, Larkin and Nagaoka \cite{Hikami80}. Even under these conditions,
that is $B/B_{tr} < 1$, the theory is unable to adequately describe the experimental data, which indicates that the current theoretical understanding of the weak localisation phenomena, in presence of spin-orbit scattering,  is incomplete.

Figure 3a shows how it is possible to fit either the low field
(central) part of the magnetoconductivity,  or the high field
tails, but not both parts simultaneously. Satisfactory fits to the high field
region can be made assuming negligible spin-orbit scattering (i.e. $\beta_{so}$ = 0). Interestingly, this fit (curve 3) yields the same dephasing rate ($\beta_{\phi}$) 
as that obtained from fitting the low field WAL peak (curves 1 and 2) where both
$\beta_{so}$ and $\beta_{\phi}$ are allowed to vary. This implies that while the dephasing
rate is field independent (as expected) the spin orbit scattering rate decreases quite
markedly as the magnetic field increases. 

If the main mechanism of spin-orbit relaxation is the spin
precession around the effective k-dependent crystal magnetic
field  that results from the bulk or structural inversion
asymmetry \cite{Dyakonov86,Lau01,Winkler03} then it would be reasonable
to assume that this precession will be affected by external
magnetic fields larger than the effective spin-orbit field
$B_{so}$. $B_{so}$ is typically smaller than 1 mT so one might
expect the spin precession to be disturbed by external fields
of this order.  We have been unable to find any theoretical
discussion of this effect in the literature and have therefore used an 
empirical approach. We postulate that the spin orbit scattering
rate decreases with increasing field in a Lorentzian fashion, ie that 
$\beta_{so} = \tau/\tau_{so} = \beta_{so}^{0}/(1+ a B^2)$. Using this expression, 
with the introduction of an additional fitting parameter {\it a}, allows quite
satisfactory fits to the experimental data (see Fig. 3(b)).
Due to space limitations discussion of the parameters
determined in the fits must be deferred to another publication but it can be
noted that the values of $a^{-1/2}$ correspond to fields of
order of 0.1 mT.

\section{Conclusion}
In contrast to intuitive expectation it is found that 
experimental weak antilocalization and
weak localization peaks respond in the same way to tilted magnetic fields
and depend only on the normal component of field. To fit the shape of the
peaks it was found necessary to assume that the
spin-orbit scattering parameter decreases with increasing 
magnetic fields. Although this can be understood qualitatively
in terms of quenching of the precession around internal, spin-
orbit related, magnetic fields an improved theoretical
treatment is needed if an understanding of the electron transport properties and
microscopical spin dynamics in gated semiconductor structures 
is to be achieved. 

\subsection{Acknowledgements}
S.A.S acknowledges support of The Canadian Institute for Advanced Research (CIAR). 
We would like to thank Chandre Dharma-wardana for helpful discussion.

\newpage

\begin{figure}[h]
\begin{center}\leavevmode
\includegraphics[width=0.5\linewidth]{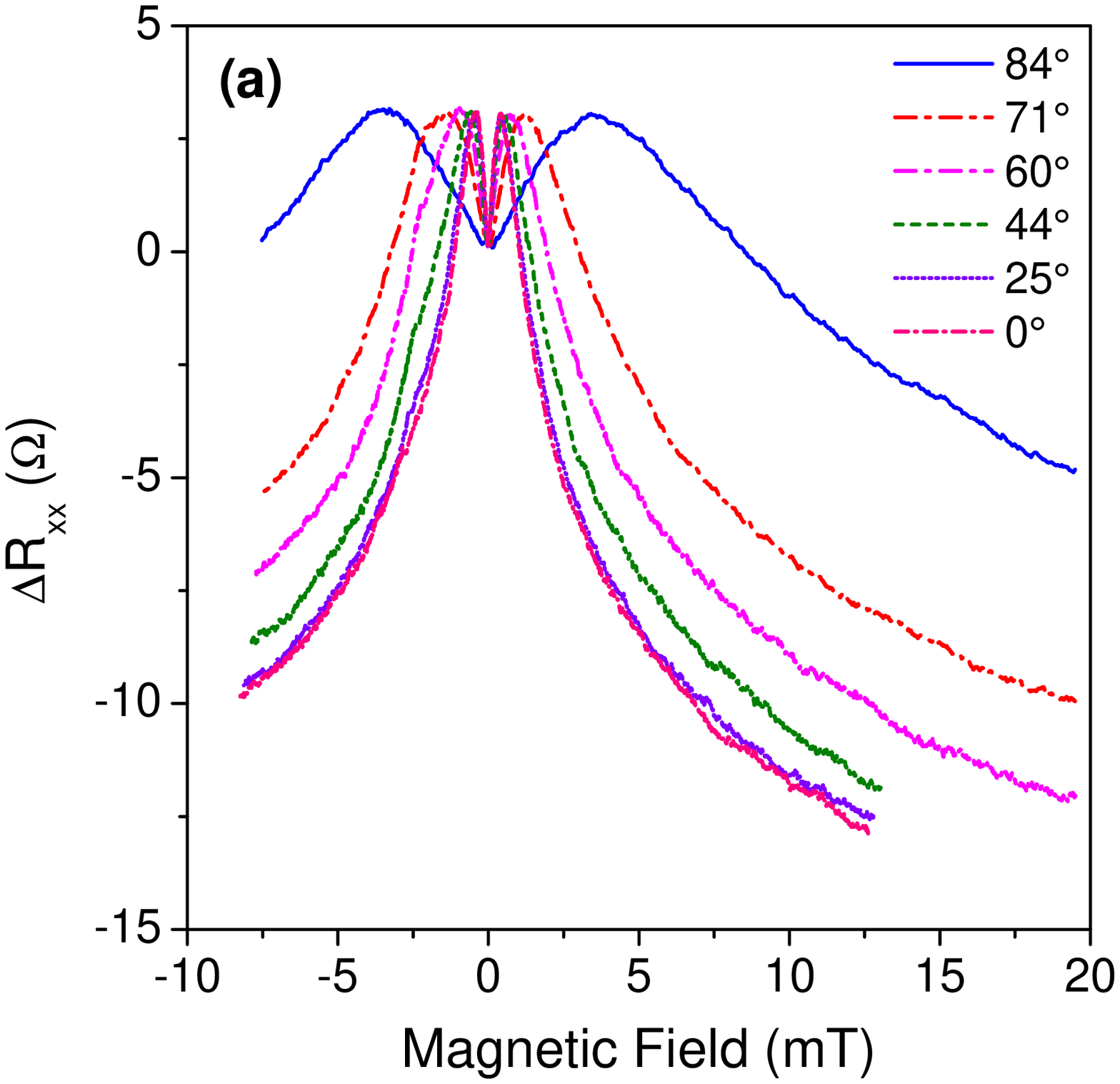}
\includegraphics[width=0.5\linewidth]{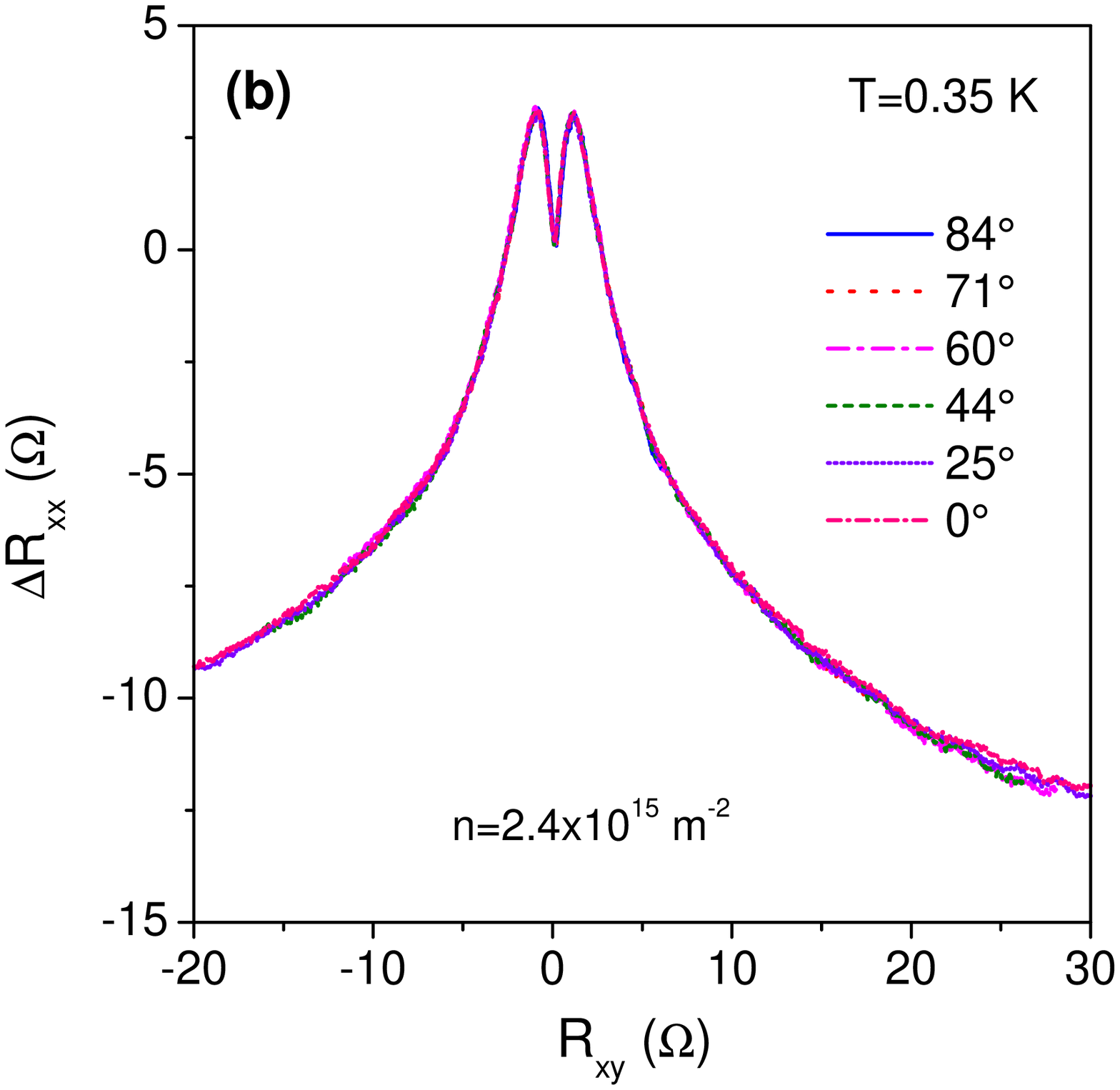}
\caption{Weak antilocalization feature in a strained
InGaAs/InP quantum well in tilted magnetic 
fields: (a) as a function of total magnetic field and (b) as a
function of the Hall resistance R$_{xy}$ which 
depends only on the perpendicular component of the field.  }
\label{fig1b} \end{center}\end{figure}

\begin{figure}[h]
\begin{center}\leavevmode
\includegraphics[width=0.5 \linewidth]{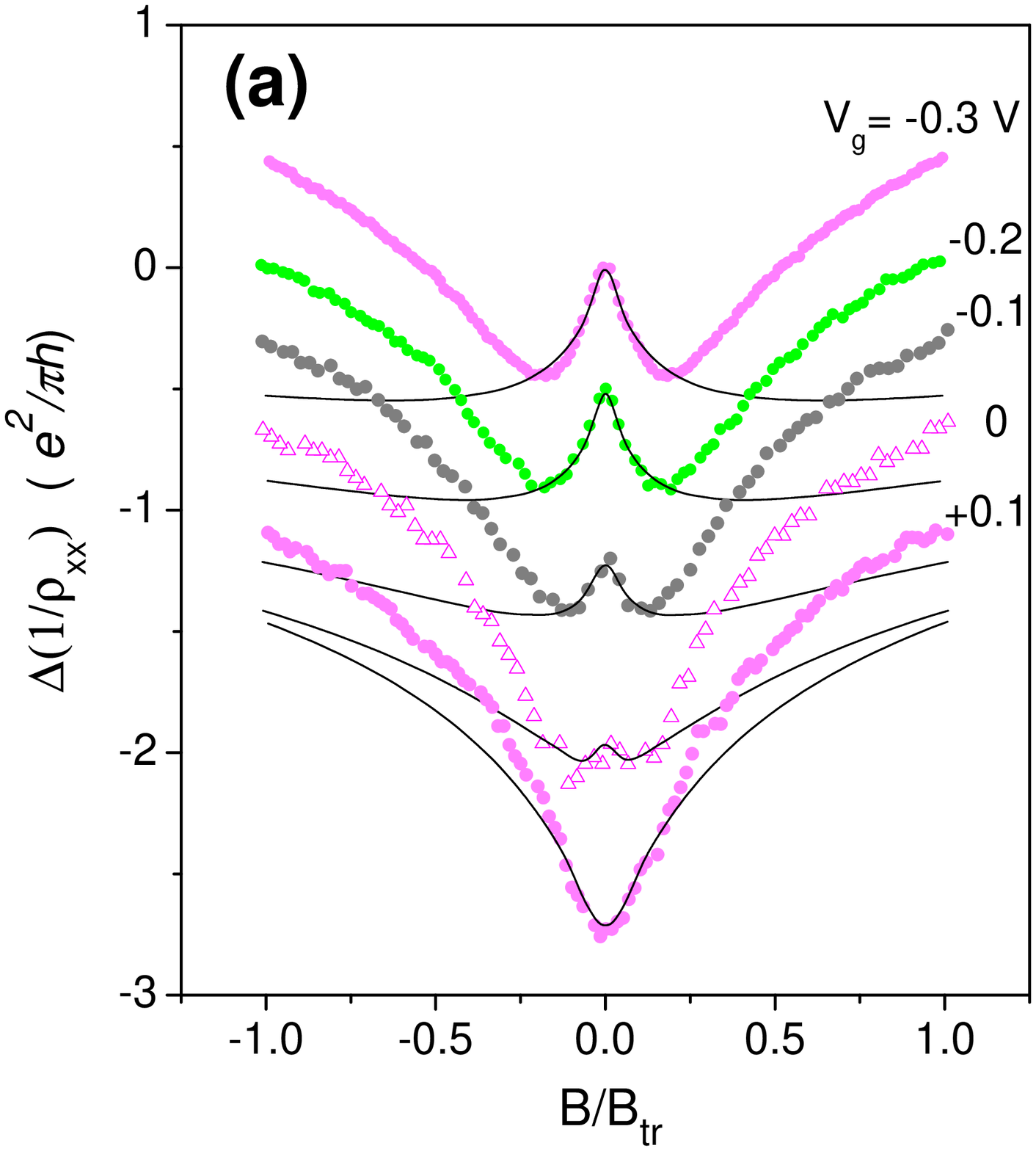}

\includegraphics[width=0.5\linewidth]{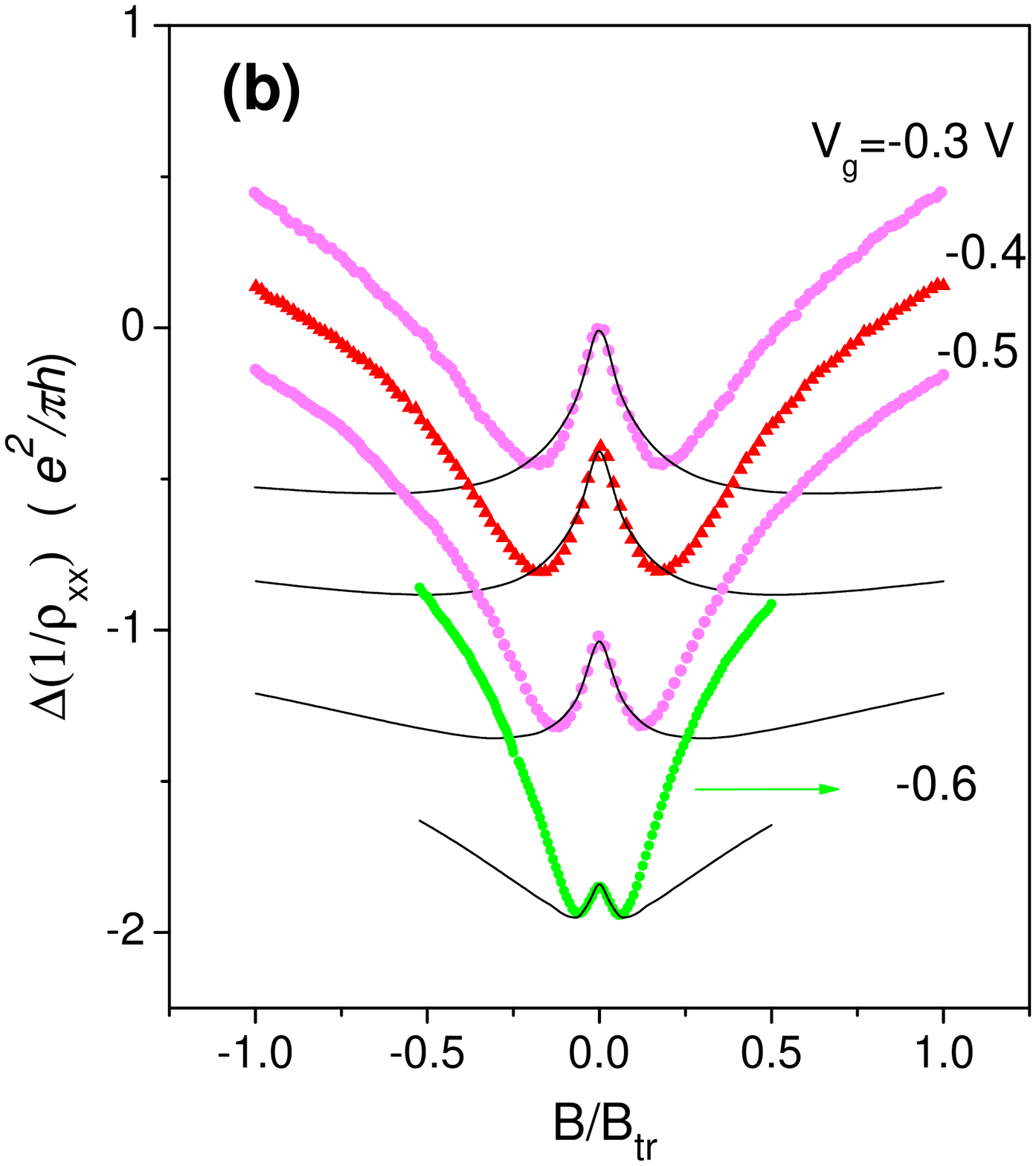}
\caption{ Experimental traces of the magnetoresistance for different gate voltages at T=0.36 K. Solid lines are best theoretical fits to the experiment using theoretical equations from \cite{Zduniak97}. }
\label{fig2b} \end{center}\end{figure}

\begin{figure}[h]
\begin{center} \leavevmode
\includegraphics[width=0.5\linewidth]{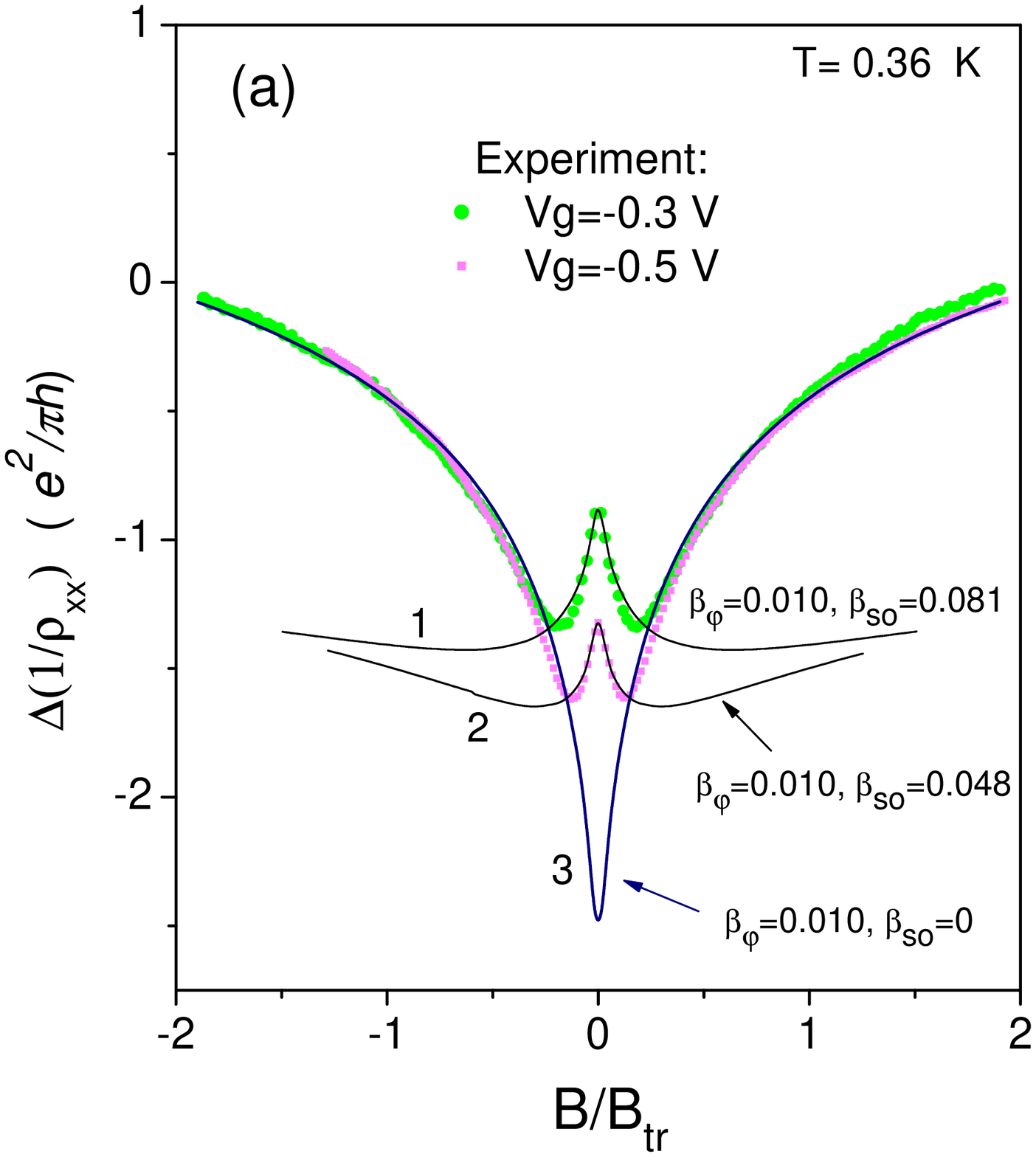}
\includegraphics[width=0.5\linewidth]{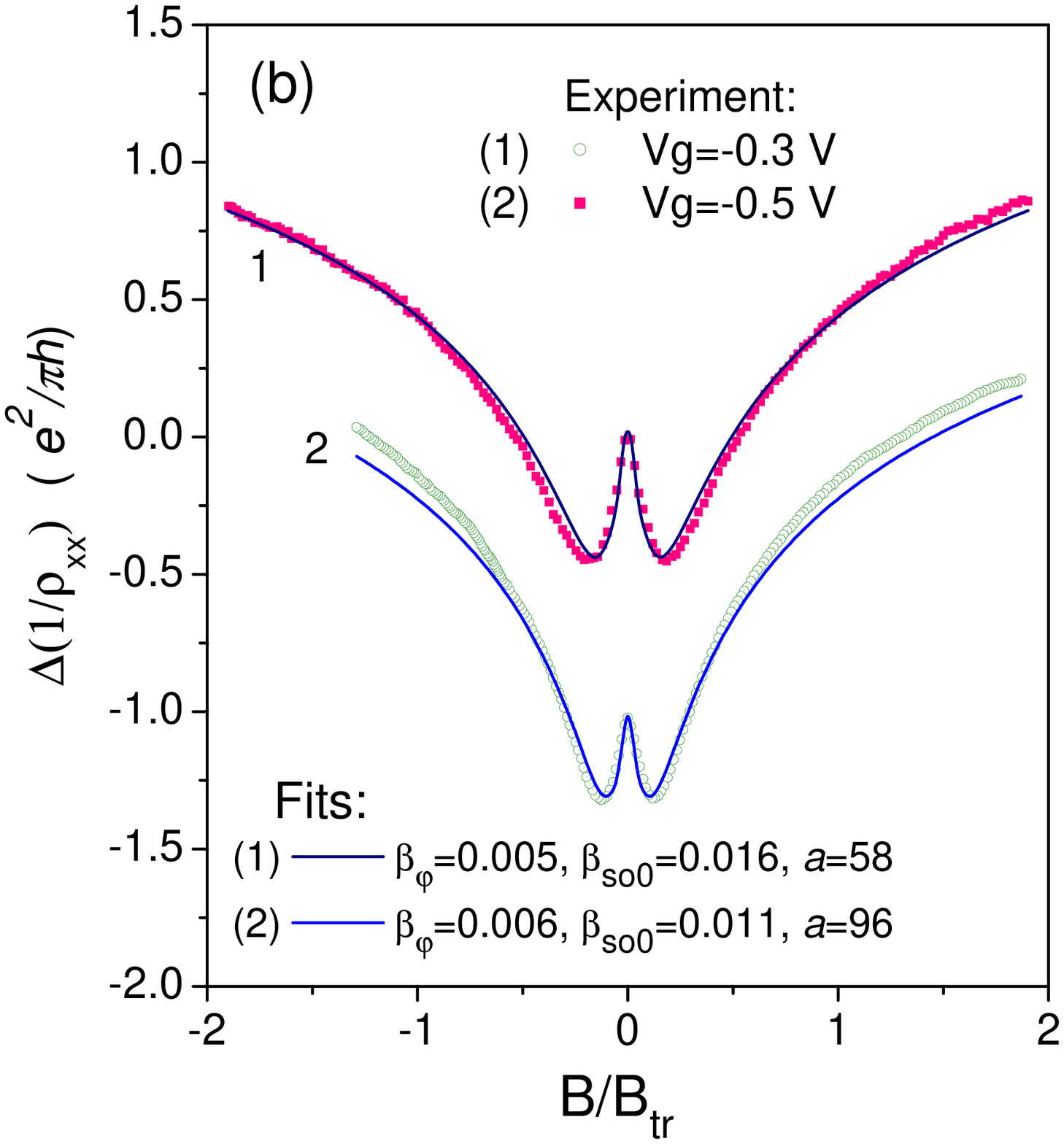}
\caption{ (a) WAL effect for two gate voltages with fitted
curves through either the central WAL peak or 
the WL tails using same phase relaxation parameter $\beta_{\phi}$ 
(b) Same experimental data as in part (a) but fitted with an
assumption that spin-orbit relaxation time 
changes with the magnetic field as 
$\beta_{so} = \tau/\tau_{so} = \beta_{so}^{0}/[1+ a (B/B_{tr})^2]$
 }
\label{fig3b} \end{center}\end{figure}

\end{document}